          [2004/06/22 1.2 (KOF); 2001/04/25 1.1 (PWD)]

\documentclass[a4paper,twocolumn]{Gaia2004} % European paper size
\usepackage{times}      % for font
\usepackage{epsfig}     % for figure inclusion
\usepackage{natbib}     % for bibliography
\title{Determination of stellar rotation with Gaia and effects of spectral mismatch}

\author[1,2]{A. Gomboc,}
\author[3]{D. Katz}
\affil[1]{University in Ljubljana, Jadranska 19, 1000 Ljubljana, Slovenia}
\affil[2]{ARI, Liverpool John Moores University, 12 Quays House, Egerton Wharf, Birkenhead, CH41 1LD, United Kingdom}

\affil[3]{Observatoire de Paris, GEPI, 5 Place Jules Janssen, F-92195 Meudon, France}

\bibpunct{(}{)}{;}{a}{}{,}  % to set bibliography punctuation to A&A style

\begin{document}


\maketitle

\begin{abstract}
  
Stellar spectra obtained with Radial Velocity Spectrometer on-board Gaia  will enable 
determination of projected rotational velocities ($v \sin i$) from rotational broadening of 
spectral lines. To estimate the accuracy with which $v \sin i$ can be determined for different 
stellar types and as a function of magnitude, we perform simulations in which "observed" 
spectra (as obtained from Gaia RVS simulator) are fitted with Kurucz spectra from template library 
using the least square method. We compare results on the $v \sin i$ accuracy as obtained in the case 
of spectra differing only in $v \sin i$ and in the case of a more general spectral mismatch (in T$_{\rm eff}$, log g, [Fe/H]).

\end{abstract}

\section{Stellar Rotation and Gaia}
Rotation of the star can have a number of important effects: as the star rotates and centrifugal forces reduce the effective gravity
according to the latitude in the star, they introduce deviations from sphericity: this has an effect on the shape of the star, its structure, luminosity and its evolution in general. 
At present, rotation is still often considered as a second order effect. Over recent years, however, a number of serious discrepancies
between current models and observations have been noticed and they show that the role of rotation has been largely overlooked.
As \citet{mae00} point out: "stellar evolution is not only a function of mass M and metallicity Z, but of angular velocity $\Omega$ as well."

To list just some of the rotation related open questions:
\begin{itemize}
\item{the influence of rotation on stellar structure and evolution, life time, position in the HR 
diagram, etc.;}
\item{rotation, mixing and abundances of He and N in massive O- and B-type stars;}
\item{relation of rotation to chemical peculiarity in stars;}
\item{the role of rotation on stellar winds and mass loss;}
\item{rotation in binary systems is connected with a number of questions, including synchronization between rotational and orbital periods, effectiveness of tidal energy dissipation, etc.;}
\item{angular momentum of stars as an indicator of massive planet's presence;}
\item{stellar rotation in open clusters as an indicator of age, binding energy etc.}
\end{itemize}
It is expected that a large statistical sample provided by Gaia will substantially contribute to solving
many of these questions related to stellar rotation. The Gaia's contribution to the field of stellar rotation
physics depends, of course, on the accuracy with which the rotational velocity will be measured, the dependence of this accuracy on stellar type and magnitude 
and consequently, on the number of stars
for which $v \sin i$ can be determined with useful accuracy. 

Radial Velocity Spectrometer (RVS) on board Gaia will enable the measurement of projected rotational velocities for a large number of stars through their spectra. The impact of the RVS characteristics, particularly its resolution, on the $v \sin i$ accuracy has been studied in \citet{gom03}. Here we present the estimation of the Gaia performances in $v \sin i$ determination based on RVS characteristics as they have been defined in December 2003.

\section{RVS Characteristics and Simulator}

\subsection{RVS Characteristics}
The RVS is a 2.0 $\times$ 1.6 degree integral field spectrograph, dispersing the light of all sources entering its field of view with resolving power $R=\lambda/\Delta \lambda=$ 11500 over the wavelength range [848,874] nm \citep{katz04b}.
The RVS will continuously and repeatedly scan the sky and a source will, on average, be observed 102 times over 5 year period.
The RVS will collect the spectra of about 100-150 million stars up to magnitude $V \simeq $ 17-18. 

\begin{figure*}[htb!]
  \begin{center}
    \leavevmode
 \centerline{\epsfig{file=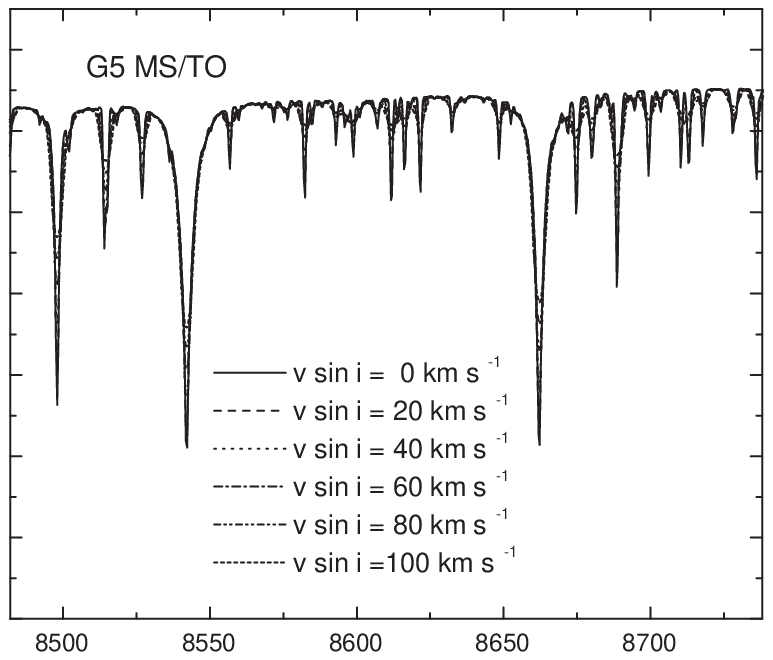, width=0.8 \linewidth}}  
 \centerline{\epsfig{file=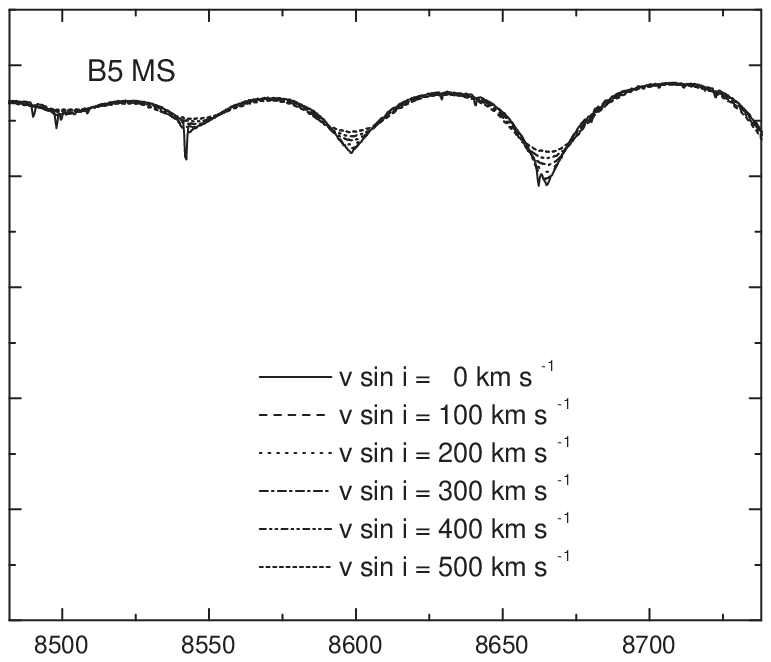,width=0.8 \linewidth}}
   \end{center}
  \caption{Examples of rotational broadening of spectral lines in Gaia RVS wavelength region for G5 MS/TO and B5 MS stars.}
  \label{rotbr}
\end{figure*}

\begin{table*}[htb!]
  \caption{Parameters of the 5 stellar types representative of Galactic population and
  considered in simulations on the accuracy of $v \sin i$ determination.}
  \label{StTy}
  \begin{center}
    \leavevmode
        \begin{tabular}[h]{lrrrrr}
         \hline \\[-5pt]
	        & T$_{\rm eff}$ [K]    &  log g      &  [Fe/H] & [$\alpha$/Fe] & $v \sin i$ [km s$^{-1}$] \\[+5pt]
	      \hline \\[-5pt]
	      K1 III  & 4500 & 2.0 & 0.0 & 0.0 & 5.0 \\
	      G5 MS/TO  & 5500 & 4.0 & 0.0 & 0.0 & 5.0 \\
	      B5 MS  & 15000 & 4.5 & 0.0 & 0.0 & 50.0 \\
	             &       &     &     &     & 150.0\\
	      K1 III  & 4500 & 2.0 & -1.5 & +0.4 & 5.0 \\
	      F5 MS/TO  & 6500 & 4.0 & -1.5 & +0.4 & 20.0 \\
	                &      &     &      &      & 50.0 \\
      \hline \\       
       \end{tabular}
  \end{center}
 \end{table*}

\subsection{Gaia RVS Simulator}
To generate stellar spectra such as will be obtained with RVS, a Gaia RVS Simulator has been developed \citep{katz04b}. The simulator uses Kurucz \citep{kur93} synthetic spectra, convolves them to the RVS spectral resolution, normalises to the star magnitude (according to its atmospheric parameters), samples and degrades them with photon, background and detector noises. The simulations take into account the zodiacal light
and faint background stars, the telescope pupil area, the instrument overall efficiency, the exposure time, the spectra profile perpendicular to the dispersion direction and the detector noise.

\begin{figure*}[htb!]
  \begin{center}
    \leavevmode
 \centerline{\epsfig{file=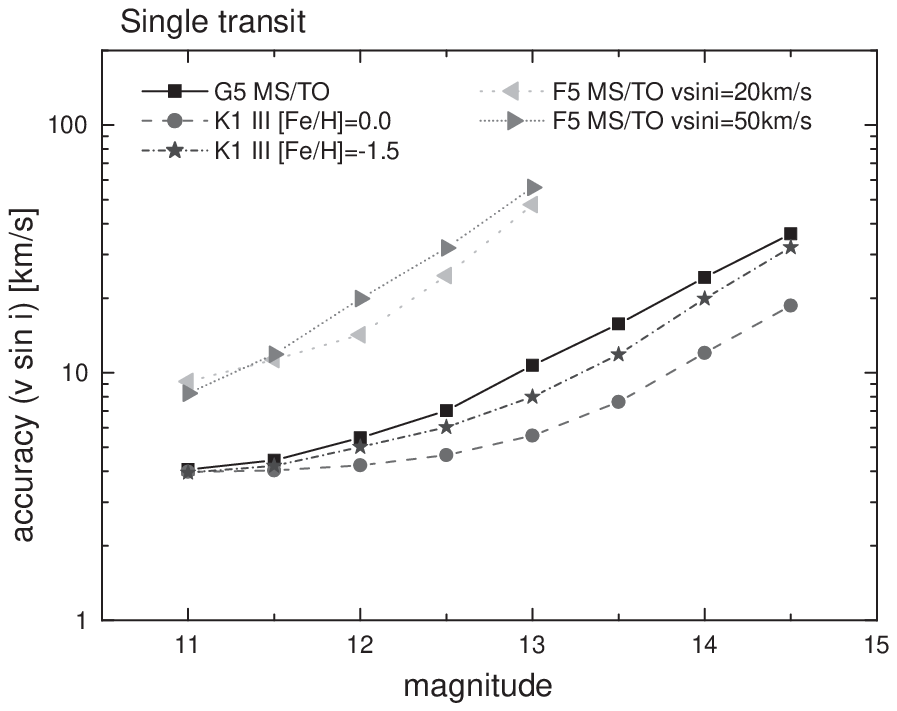, width=0.52\linewidth} \quad \epsfig{file=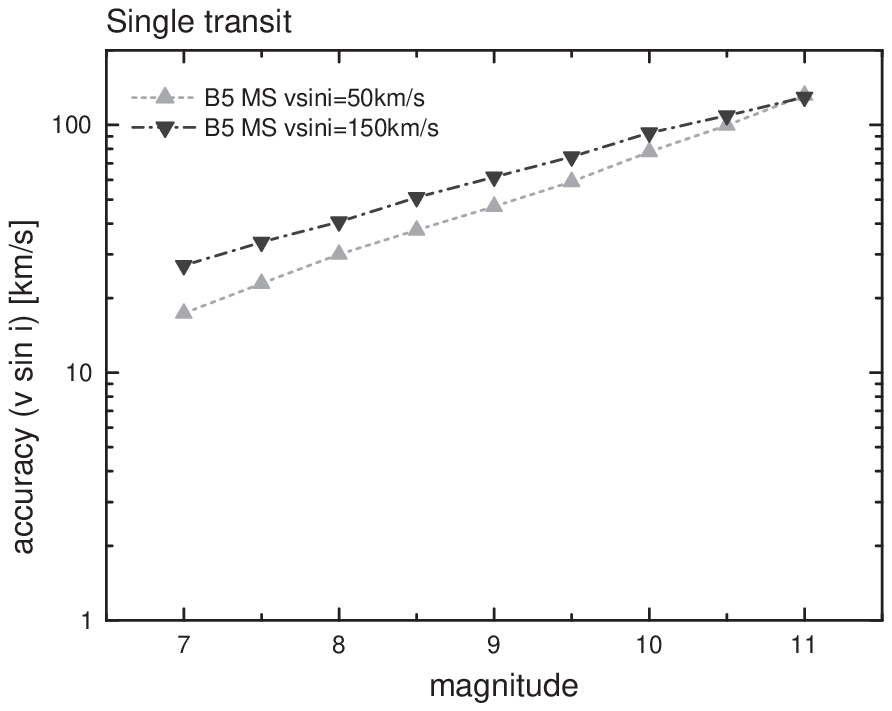,width=0.52\linewidth}}
  \centerline{\epsfig{file=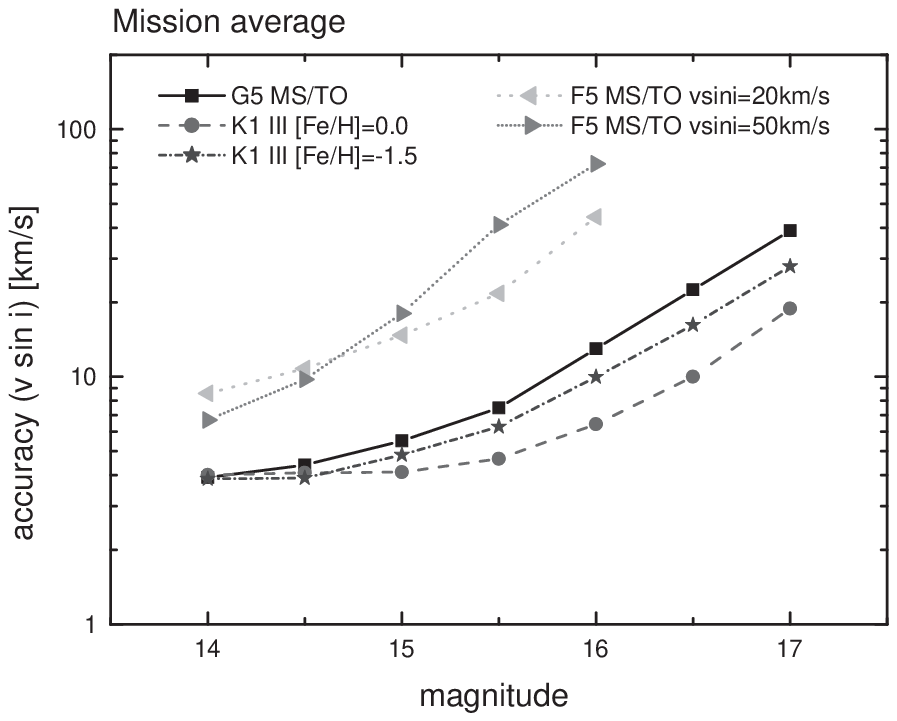, width=0.52\linewidth} \quad \epsfig{file=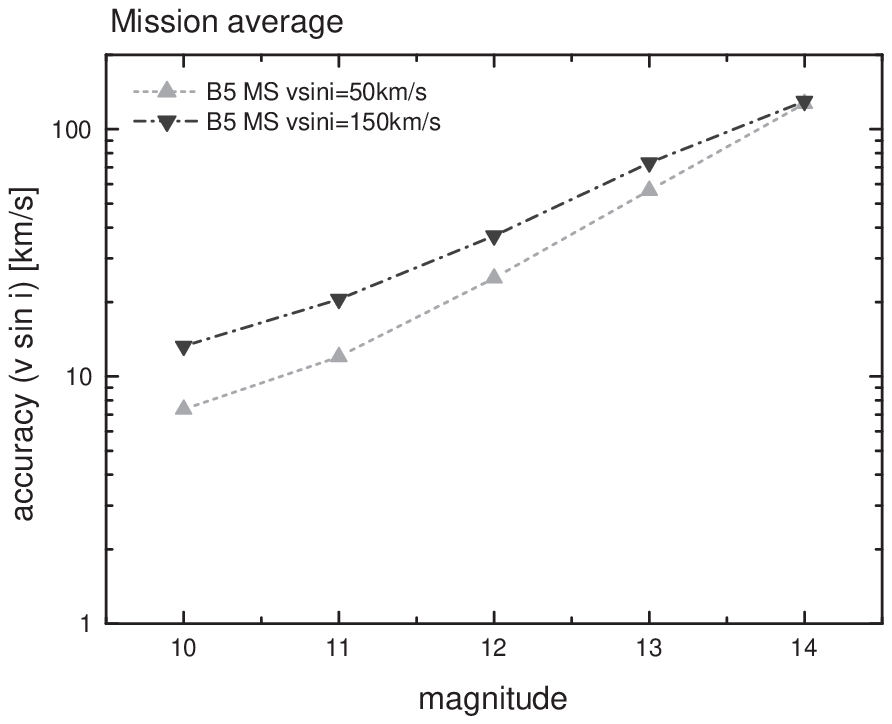,width=0.52\linewidth}}
 \end{center}
  \caption{Accuracy of $v \sin i$ for a single stellar transit over the RVS focal plane (top) and at the end of the mission
  (bottom) as a function of magnitude and for stellar types from Table \ref{StTy}.}
  \label{vsini}
\end{figure*}

\section{Simulations and RVS Performances on $v \sin i$}
Analysis of stellar spectra obtained with RVS will, in addition to radial velocity, also enable the determination of other stellar
parameters, including the projected rotational velocity ($v \sin i$)
of individual stars. The spectroscopic determination of $v \sin i$ is based on rotational broadening of 
spectral lines, illustrated in Figure \ref{rotbr}.

Among analysis methods commonly used to determine $v \sin i$, we chose the simple least square fit method.
The simulations are performed with synthetic spectra for single stars in following way:
(i) an "observed" spectrum is obtained with Gaia-RVS simulator, (ii) fitted with "template" noise-free spectra with different $v \sin i$ 
from the library and (iii) the template with minimum square deviations is taken as the best fit.
This procedure is repeated for N=1000 spectra and the accuracy is defined as the standard 
deviation error between original and recovered $v \sin i$:  
\begin{equation}
\sigma^2 = {1\over N}\sum_{k=1}^N \Bigl(v \sin i_{k}^{\rm orig} - v \sin i_{k}^{\rm rec}\Bigr)^2
\label{sigma}
\end{equation}

We considered 5 stellar types (Table \ref{StTy}), which are representative tracers of the 
Galactic structure. Library step size in $v \sin i$ was 1 km s$^{-1}$, 2 km s$^{-1}$ and 5 km s$^{-1}$
for K1 III and G5 MS/TO, F5 MS/TO and B5 MS/TO stars, respectively.

Results are presented in Figure \ref{vsini} for a single transit of the star over the RVS focal plane
and at the end of the mission (after 102 transits of a star on average): 

- in a {\bf single transit}  
\begin{itemize}
\item{for late type stars the accuracy of up to about 10 km s$^{-1}$
should be achievable to the magnitude of 13 - 14;}
\item{for F5 MS/TO stars the accuracy of about 20 km s$^{-1}$ is expected to the magnitude of about 12;}
\item{for fast rotators such as B5 MS star, the accuracy of several 10 km s$^{-1}$ should be 
obtained for very bright, 7 - 8 magnitude stars.}
\end{itemize}
This difference for B5 MS star is understandable in view of the fact that spectral lines in these stars
are very wide and rotational broadening is far from dominating the line profile and its width (Figure \ref{rotbr}).

- at the {\bf end of the mission} the accuracy shows similar behaviour, but shifted towards fainter magnitudes
for about 2 - 3 magnitudes:
\begin{itemize}
\item{for late type stars accuracy of up to about 10 km s$^{-1}$ is expected for magnitudes about 16;}
\item{for F5 MS/TO stars the accuracy of about 20 km s$^{-1}$ should be obtained up to the 
magnitude of about 15;}
\item{for B5 MS stars the accuracy of the order of 10 km s$^{-1}$ could be achieved for stars with 11 - 12 magnitude.}
\end{itemize}
We note that there is a bias in these results which is most evident for bright magnitudes and is of the order of few km s$^{-1}$. It is introduced due to varying position of the start of the spectrum on the first CCD pixel. The interpolation of template spectra to same wavelength bins introduces slight variations in the line profiles, which mimic the rotational broadening. A consequence is that the fitting procedure favours lower $v \sin i$ templates as the best fit.

 \begin{figure*}[htb!]
  \begin{center}
    \leavevmode
 \centerline{\epsfig{file=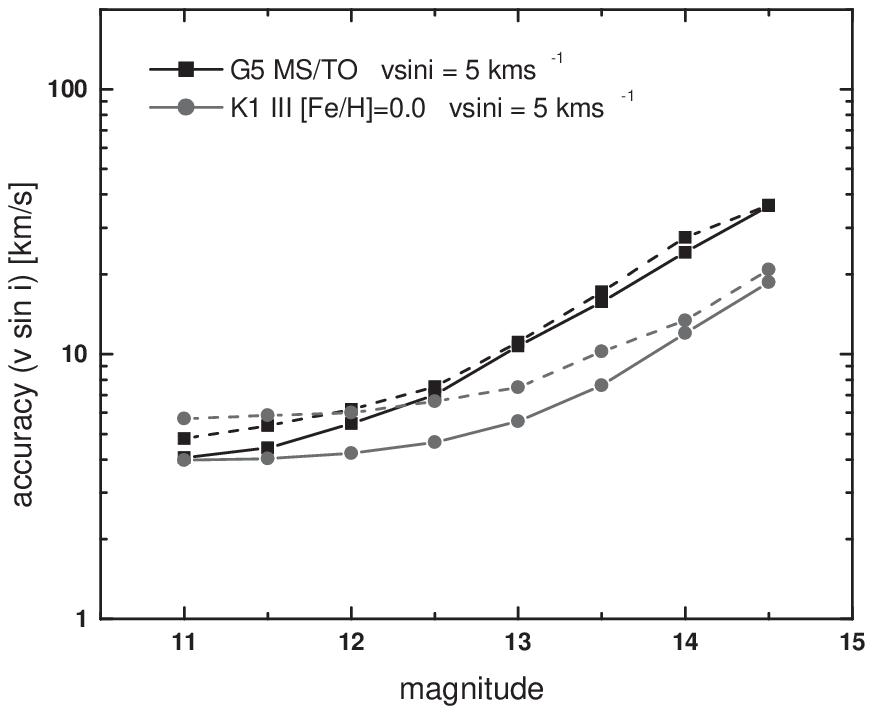, width=0.5\linewidth} \quad \epsfig{file=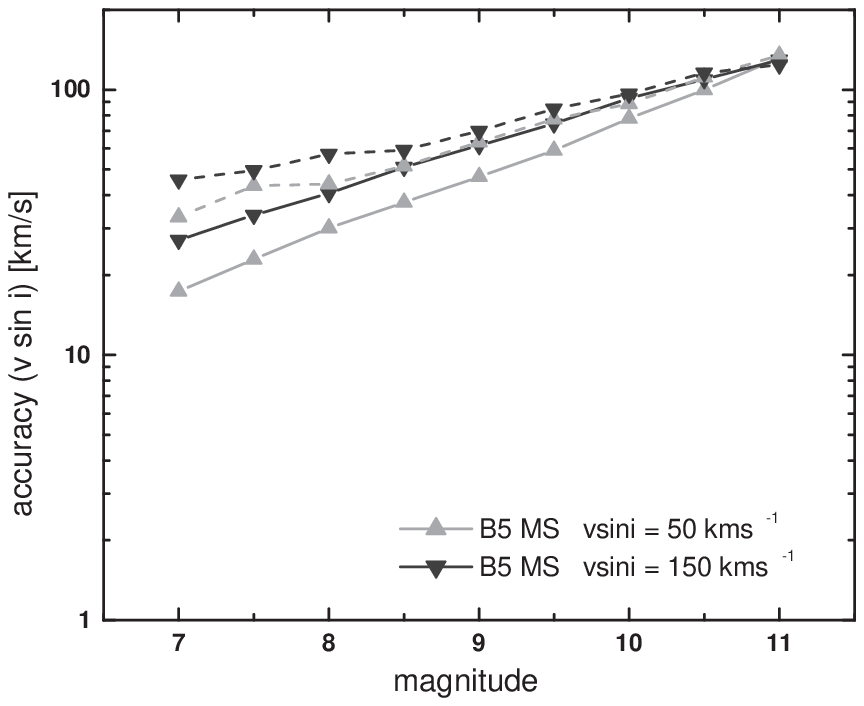,width=0.5\linewidth}}
   \end{center}
  \caption{Accuracy of $v \sin i$ in the case of no spectral mismatch (solid lines) and 
  in the case of mismatch in effective temperature (dashed lines). Assumed uncertainty
  is $\Delta$ T$_{\rm eff}$ = 150~K for late type stars (left) and $\Delta$ T$_{\rm eff}$ = 500~K for 
  B5 MS stars (right).}
  \label{vsiniT}
\end{figure*}

 \begin{figure*}[htb!]
  \begin{center}
    \leavevmode
 \centerline{\epsfig{file=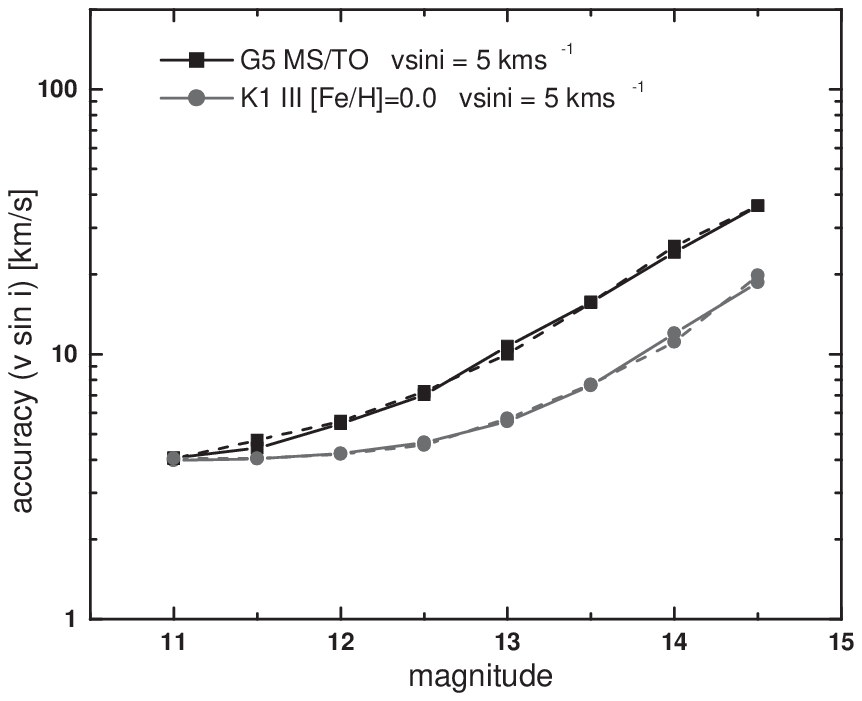, width=0.5\linewidth} \quad \epsfig{file=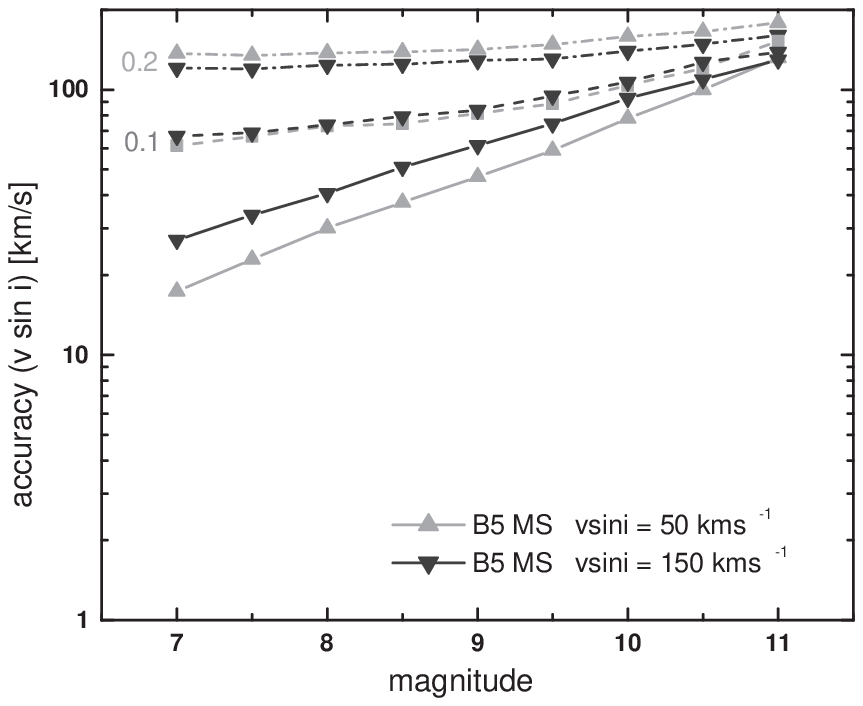,width=0.5\linewidth}}
   \end{center}
  \caption{Accuracy of $v \sin i$ in the case of no spectral mismatch (solid lines) and 
  in the case of mismatch in gravity (dashed lines). Assumed uncertainty
  is $\Delta$ log g = 0.2 for late type stars (left) and $\Delta$ log g = 0.1 and 0.2 for 
  B5 MS stars (right).}
  \label{vsinig}
\end{figure*}

 \begin{figure*}[htb!]
  \begin{center}
    \leavevmode
 \centerline{\epsfig{file=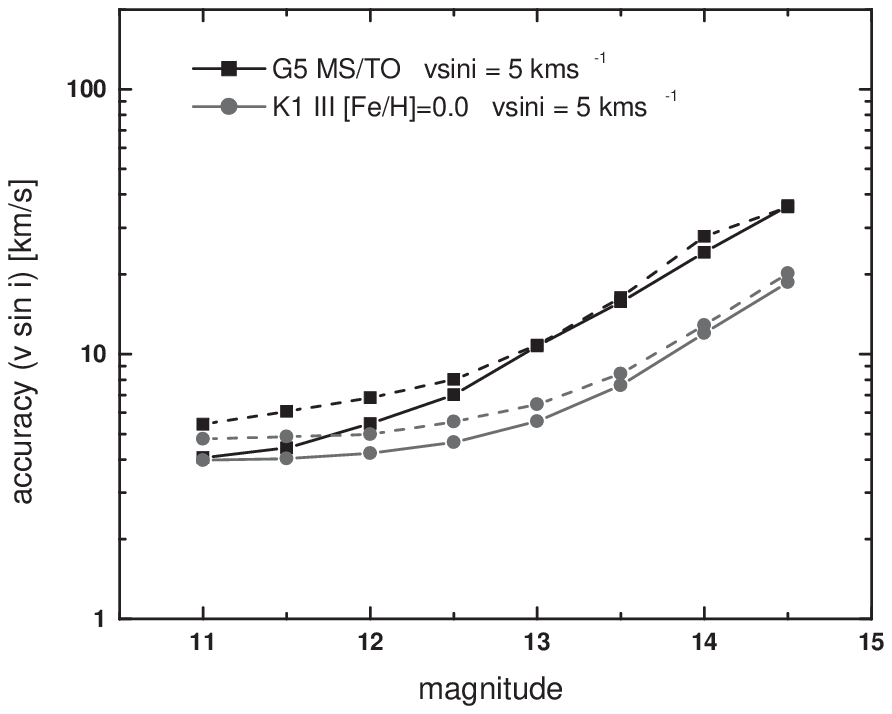, width=0.5\linewidth} \quad \epsfig{file=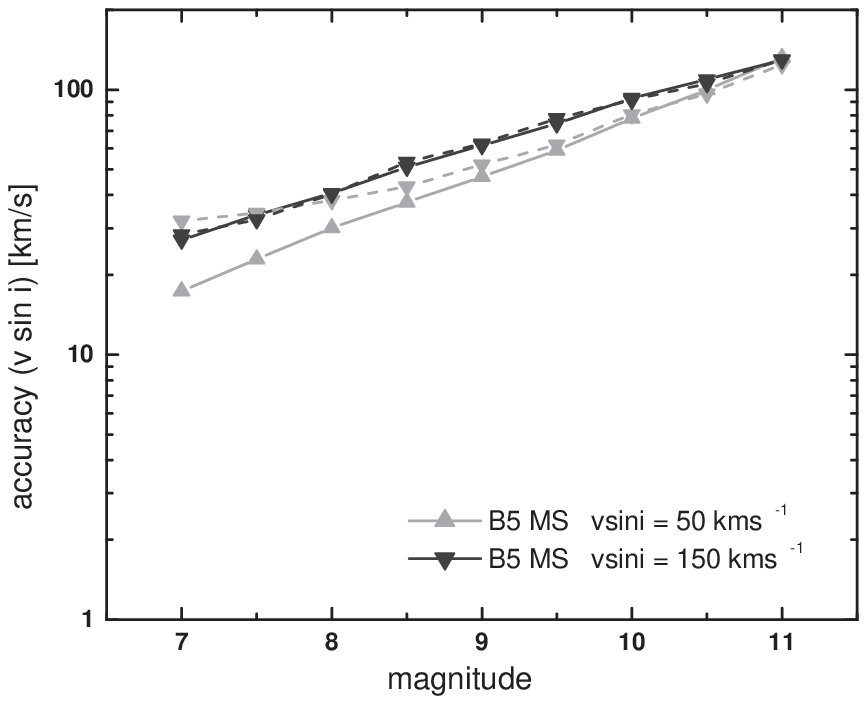,width=0.5\linewidth}}
   \end{center}
  \caption{Accuracy of $v \sin i$ in the case of no spectral mismatch (solid lines) and 
  in the case of mismatch in metallicity (dashed lines). Assumed uncertainty
  is $\Delta$ [Fe/H] = 0.2 for late type stars (left) and $\Delta$ [Fe/H] = 1.0 for 
  B5 MS stars (right).}
  \label{vsinim}
\end{figure*}

Results in Figure \ref{vsini} together with statistical numbers of stellar population indicate that Gaia should be able to measure with useful accuracy projected rotational 
velocities of about 25 - 50 million stars.
We would like to note, though, that these results are quite optimistic in a sense, that we have not 
included a number of effects, and therefore should be considered only as an estimate. 

Among the effects which may aggravate the accuracy of $v \sin i$ determination and were not taken into account, are: 
\begin{itemize}
\item{Modeling uncertainties - all simulations presented here were performed with synthetic spectra 
and therefore any discrepancies
between spectra given by stellar models and observed spectra are not taken into account. One of the 
possible improvements in this area is more accurate inclusion of rotation in spectra modelling (see \citet{jau05}).}
\item{Crowding - as Gaia RVS is a slitless spectrograph the stellar spectra will increasingly overlap with decreasing Galactic latitude  and  increasing 
number of stars in the field. As studied by \citet{zwi03} and \citet{katz04a} it will be possible to recover most of the overlapped spectra with the help of
astrometric and photometric information from other instruments onboard Gaia. It is expected though, 
that details in the line profile, such as accurate rotational broadening, might be lost to some extent.}
\item{Spectral mismatch - results presented in Figure \ref{vsini} and discussed so far, are based on comparison of an
original "observed" spectrum with library spectra differing {\it only} in $v \sin i$. However, determinations of other stellar parameters will
have their own uncertainties, which may affect the $v \sin i$ determination.}
\end{itemize}

\section{Effects of Spectral Mismatch on $v \sin i$ accuracy}
To include the effects of spectral mismatch on the accuracy of obtained $v \sin i$ we repeated simulations, described
in previous section, for a single transit across the RVS focal plane, but took into account uncertainties in effective temperature, gravity and metallicity of the star. We assumed a Gaussian
distribution of individual parameters with standard deviations equal to uncertainties as predicted by \citet{van03}. 

Results show that effect of the mismatch in effective temperature (Figure \ref{vsiniT}) is not substantial.
Similarly it holds for the log g mismatch for late type stars (Figure \ref{vsinig} left). 
However, log g mismatch for B5 MS seems to introduce a large aggravation in $v \sin i$ accuracy
(Figure \ref{vsinig} right). Finally, the mismatch in metallicity seems not to be crucial for $v \sin i$ determination (Figure \ref{vsinim}).

These results look quite promising, since they seem to indicate that in general, accuracy of $v \sin i$ 
will not suffer greatly for expected uncertainties in other stellar parameters.
However, the effect of combination of these uncertainties on $v \sin i$ determination 
still remains to be studied.

\section{Conclusion}
Simulations presented here indicate that the accuracy of projected rotational velocity obtained by 
RVS instrument onboard Gaia will be of about 10 km s$^{-1}$ for late type stars of magnitude 13 - 14
in a single transit and of magnitude 16 at the end of the mission. For fast rotating B5 MS stars, 
the accuracy of several 10 km s$^{-1}$ is expected for 7 - 8 magnitude, in a single transit, 
and 11 - 12 magnitude stars at the end of the mission. 

We also investigated the influence of 
spectral mismatch and results indicate
that uncertainties in effective temperature, gravity and metallicity, as expected from Gaia,
will not substantially influence the $v \sin i$ accuracy in general. Future work, including study 
of combined effect of these uncertainties and the effect of spectra overlapping in crowded Galactic 
regions, is needed.

From current results presented here we estimate that Gaia can be expected to be able to determine 
with useful accuracy the $v \sin i$ for about 25 to 50 million stars.
Since this will exceed the current number of stars with measured $v \sin i$ \citep{gle00} by a factor of thousand,
it will undoubtedly greatly contribute to our knowledge on stellar rotation.

\section*{Acknowledgments}
AG acknowledges financial support of the Slovenian Ministry of Science, Education and Sport
and the receipt of the Marie Curie Fellowship from the European Commission.
DK acknowledges financial support from CNES.

\end{document}